\documentclass[a4paper,10pt]{article}

\usepackage[utf8]{inputenc}

\def\bgphi{{\mbox{\boldmath $\Phi$}}}
\def\bgTheta{{\mbox{\boldmath $ \Theta$}}}

\title{  Topological Defects in   a  Deformed  Gauge  Theory   }
\author{ Mir Faizal$^{1,2}$ and   Tsou Sheung Tsun$^{3}$  \\ \\
\\
$^1$Irving K. Barber School of Arts and Sciences, \\ University of British Columbia - Okanagan,
\\ Kelowna, British Columbia V1V 1V7, Canada
\\\\$^2$Department of Physics and Astronomy,\\
University of Lethbridge, \\
Lethbridge, Alberta, T1K 3M4, Canada
\\ \\
$^3$Mathematical Institute, University of Oxford,\\ Andrew Wiles
Building, \\Radcliffe Observatory Quarter, Woodstock Road,
\\ Oxford
OX2 6GG, United Kingdom }
\date{}

\begin{document}

\maketitle

\begin{abstract}
In this paper, we will analyse the topological defects in a  deformation of a non-abelian gauge theory using the 
Polyakov variables. The gauge theory will be deformed by the existence of a
 minimum measurable length scale in the background spacetime. We will construct the Polyakov loops for this deformed 
 non-abelian gauge theory, and use these deformed   loop space variables for obtaining a deformed 
  loop space curvature. It will be demonstrated that this curvature will vanish if the deformed Bianchi identities are satisfied. 
  However, it is possible that the original Bianchi identities are satisfied, but the deformed Bianchi identities are 
  violated at the leading order in the deformation parameter, due to some topological defects. Thus, 
 topological defects could be produced purely from a deformation 
  of the background geometry. 
\end{abstract}
\section{Introduction}

Topological defects can be analysed using the Polyakov variables, and these Polyakov variables are defined using the  
   the holonomies of the gauge fields \cite{p1, t1, t2, p}.  In this paper, we will call these 
holonomies of the gauge fields as Polyakov loops, as they were introduced by Polyakov \cite{p1}. 
 So, these Polyakov loops  would be  constructed using the gauge fields  as   
 the holonomies of closed
loops in spacetime. They are also called as the 
Dirac phase factors in the 
  physics literature. They do not depend on the
  parameterization
chosen, and  they capture some interesting topological properties of the gauge theory. 
In fact, they resemble the Wilson's loops, but 
  unlike the Wilson loops, no trace is taken over the gauge group for Polyakov loops. 
  Thus, in this paper, there is a  difference  between these Polyakov loops, and 
  the usual   Wilson's loops. This is because Wilson's loops  only represented by a number, 
  but Polyakov loops are   gauge group-valued functions of the
infinite-dimensional loop space \cite{p1}.   So, they can be used to analyse  various interesting  structures in the gauge theory, 
and this includes topological defects produced by the existence of non-abelian monopoles. It may be noted that recently 
Polyakov loops have been used for analyzing various interesting physical systems including fractional M2-branes \cite{m2frac}. 
They have also been used for analyzing   three dimensional supersymmetric 
gauge theory \cite{pqaa}, and these theories are important to study systems like M2-branes and D2-branes.
 The four dimensional supersymmetric gauge theories have also been analysed  using this formalism 
\cite{pq1a}. In fact, this formalism was used to analyse the non-abelian monopoles in four dimensional 
supersymmetric gauge theories. 
Thus, it is possible to use this formalism to analysing various interesting generalizations of the usual gauge theories.
So, in this paper, we will analyse the effect of topological defects on a deformed non-abelian gauge theory. This gauge theory 
will be deformed by the existence of a minimum measurable length scale in the background geometry. 

Such a deformation of the gauge theories by the existence of a minimum measurable length scale in the background geometry is in turn 
motivated from  low energy effects of   quantum gravity. This is because almost all the approaches to quantum gravity 
restrict the measurement of spacetime below    the Planck scale. The string theory is one of the most important approaches  
for analyzing quantum   gravity,
and   the  fundamental string is the smallest probe  available in perturbative string theory, 
and so it is not possible to probe spacetime
below the string length scale in string theory   \cite{z2,zasaqsw,csdcas,cscds,2z}. 
{  Thus, the   string length acts, which is given by $l_s = \alpha'$ as   a minimum measurable length in string theory.
Furthermore,  if  non-perturbative effects are taken into consideration, then it is possible to have $D0$-brane which is a point 
like object. However, it has been argued that even in presence of such brane, there is an intrinsic minimal length of the order of 
$l_{min} = l_s g_s^{1/3}$, where $g_s$ is the string coupling constant  \cite{s16, s18}. 
  The total energy of the quantized string depends on the  excitation $n$ and winding number $w$, and under T-duality the $n$ and $w$ 
  gets interchanged, as $R \to l_s^2/R$ and $n \to w$. So, a description of string theory below $l_s$ is the same as the 
  description above it, and so it can be argued from T-duality that the 
  string theory cannot be described below the string length scale  \cite{s16}. 
  The T-duality has  been used to construct an   effective path integral  for  
 the center of mass of the string, and analyze the  corresponding Green’s function  \cite{green1, green2}. 
 This has been done by analyzing strings propagating in spacetime with compactified additional dimensions. 
 It has been demonstrated that this Green's function  also has  an minimal length associated with it \cite{green1, green2}.   
 So, string theory due to T-duality has a minimal length associated with it. It may be noted that this minimal length 
 can be different from Planck length \cite{s16}. This is because  the the Planck length $l_{PL}$ can be expressed 
as $l_{PL} = g_s^{1/4} l_s$   \cite{s16}.  
It  has also been argued that a minimal length may  exist in models of quantum gravity, such as the loop quantum 
gravity   \cite{z1}. The 
  physics of black holes restricts the measurement to scales larger than the Planck scale.  This is because the 
energy needed to probe a region of spacetime below Planck scale is greater than the energy needed to 
form a mini black hole in the region of spacetime \cite{z4,z5}. So, if we try to probe the spacetime at a scale smaller than the
Planck scale, a mini black 
hole will form in that region of spacetime, and this will in turn restrict our ability to analyse that region of spacetime. 
It has also been argued that this length can be much larger than the Planck length, and its scale would be fixed by 
present experimental data \cite{ml12, ml14}.  So, it may be possible to have such effects observed in future experiments, 
and thus it would be interesting to study different aspects of such effects. }

 However, the problem with the existence of such  a minimum measurable length scale is that it is not 
  consistent with the foundations principles of 
ordinary quantum mechanics. This is because the ordinary quantum mechanics is based on the 
the  Heisenberg uncertainty principle, and according to this principle  it is   possible to detect the position of 
a particle with arbitrary accuracy, if  the  momentum   is not measured. 
Thus,  according to the Heisenberg uncertainty principle there is no bound on the accuracy to 
which the length can be measured as long as the momentum is not measured. 
Thus, in principle, according to the Heisenberg uncertainty principle,  we can analyse the spacetime at a length scale smaller 
than the Planck scale, and so no minimum measurable length scale exists.
However, the  Heisenberg uncertainty principle  can be modified to incorporate the existence of  
  a minimum measurable length scale. This can be done by deforming the usual 
Heisenberg uncertainty principle  $  \Delta x \Delta p \geq  {1}/{2}  ,  
$ to 
$  \Delta x \Delta p \geq {1}/{2} (1 + \beta (\Delta p)^2), 
$ where $\beta$ is a parameter in the theory.
 This modified Heisenberg uncertainty principle  is called the 
  generalized uncertainty principle (GUP).

As the Heisenberg
uncertainty principle is closely related to the Heisenberg algebra, such a deformation of the 
Heisenberg uncertainty principle will generate a deformation of the  Heisenberg algebra. In this   GUP deformed Heisenberg algebra, 
the  commutator of momentum and position  operators is a function of 
  momentum,  $
 [x_i, p_j] = i   (\delta_{ij} + \beta (p^2 \delta_{ij} + 2 p_i p_j))$ \cite{2z,14,17,18,51, 55}. 
This deformation of the Heisenberg algebra will also produce a deformation of the  coordinate representation 
of the momentum operator. In fact, it is possible to write the deformed  momentum operator,   to the first order in $\beta$, as  
$
 p_i  = -i  \partial_i (1 - \beta   \partial^j \partial_j) 
$.   In this paper, we will analyse a relativistic version of this deformation, and 
the corresponding gauge theory using the Polyakov loop formalism. 
\section{Loop Space}
In this section, we will construct the Polyakov loops for a deformed gauge theory, which will be deformed by the 
deformation of the Heisenberg algebra by GUP. 
 It is also possible to define a relativistic version of the GUP deformed Heisenberg algebra, 
and study the  quantum field theory corresponding to such a deformed algebra \cite{qft1, qft2,qft4,qft5,qft6,qft7}. 
Thus, the full covariant algebra can be written as, 
$
\left[\hat x^\mu,\hat p_\nu\right]=i\delta^\mu_\nu \left[1+\beta \hat p^\rho \hat p_\rho\right]+2i\beta \hat p^\mu \hat p_\nu.
\label{generalized_uncertainty_principle}
$
The generalized uncertainty   for this deformed algebra can be expressed as $
 \Delta x^\mu \Delta p_\mu \geq  {1}/{2}\left(1+3\beta \Delta p^\mu \Delta p_\mu
+3\beta \langle p^\rho \rangle \langle p_\rho \rangle\right), $ and this generalized uncertainty can be used to obtain the following bound 
$
 \Delta x^\mu_{min}=\sqrt{3\beta}\sqrt{1+3\beta\langle p^\rho \rangle \langle p_\rho \rangle} 
$ \cite{qft1}.
So, there exists a  minimum length $l_s$ and a minimum time $t_s$ in this algebra, such that  
$
l_s=\sqrt{3\beta}, $ and $ \quad t_s=\sqrt{3\beta}. 
$
To the first order in $\beta$, we can write the deformed momentum operator as  
$
p_\mu= -i \partial_\mu ( 1 - \beta \partial^\rho \partial_\rho )+\mathcal{O}\left(\beta^2\right). 
$
It is possible to define a gauge covariant derivative which is consistent with the existence of a minimum length scale 
as 
\begin{equation}
\mathcal{D}_\mu=\left(1-\beta D^{\rho}D_\rho\right)D_\mu,
\label{covar} 
\end{equation}
where 
$
 D_\mu = \partial_\mu +i A_\mu^AT_A.
$
Here $T_A$ are the generators of the Lie algebra 
$ [T_A, T_B] = i f_{AB}^C T_C. $ 
Now as the covariant derivative transform  
$
 D_\mu \to U D_\mu U^{-1}, 
$
so the deformed covariant derivative transform as \cite{qft2}
\begin{eqnarray}
\mathcal{D}_\mu &\to&  -i\left(1-\beta  UD^{\rho} U^{-1} UD_\rho
  U^{-1}\right)U D_\mu U^{-1} \nonumber\\
  &=&-i U \left(1-\beta   D^{\rho} D_\rho \right)  D_\mu U^{-1} \nonumber\\
 &=& U\mathcal{D}_\mu U^{-1}.
\end{eqnarray} 
So, the GUP deformed covariant derivative still transforms like a regular covariant derivative. 
It is possible to show that the Bianchi identity will hold, but the
algebraic manipulations are long  (we used the  package Quantum Mathematica to prove this result),  
\begin{eqnarray}
  && [  \mathcal{D}_\lambda, [  \mathcal{D}_\mu,   \mathcal{D}_\nu]]  +[  \mathcal{D}_\mu, [  \mathcal{D}_\nu,   \mathcal{D}_\lambda]]  
 +[   \mathcal{D}_\nu, [  \mathcal{D}_\lambda,   \mathcal{D}_\mu]]  
\nonumber \\   &=& 
 [ (1 - \beta D^\rho D_\rho)  {D}_\lambda, [ (1 - \beta D^\tau D_\tau)   {D}_\mu,  (1 - \beta D^\sigma D_\sigma)  {D}_\nu]] 
 \nonumber \\ && +[ (1 - \beta D^\tau D_\tau)  {D}_\mu, [ (1 - \beta D^\sigma D_\sigma)   {D}_\nu,  (1 - \beta D^\rho D_\rho)   {D}_\lambda]]  
\nonumber \\ &&  +[  (1 - \beta D^\sigma D_\sigma)  {D}_\nu, [ (1 - \beta D^\rho D_\rho)  {D}_\lambda, (1 - \beta D^\tau D_\tau)   {D}_\mu]]  
 \nonumber \\&=&  0.
\label{bianchi}  
\end{eqnarray}
Motivated  from the definition of the usual field tensor  $ {F}_{\mu\nu}=-i[ D_\mu, D_\nu]$,  the deformed   field tensor is defined as, 
\begin{eqnarray}
\mathcal{F}_{\mu\nu}&=&-i\left[\mathcal{D}_\mu,\mathcal{D}_\nu\right]
\nonumber \\ &=&-i\left[\left(1-\beta D^\rho D_\rho\right)D_\mu,\left(1-\beta D^\rho D_\rho\right)D_\nu\right]\nonumber\\
&=&\left(1-\beta D^\rho D_\rho\right)\left[\left(1-\beta D^\rho D_\rho\right)F_{\mu\nu}
  -\beta\left(D^\rho F_{\mu\rho}D_\nu-D^\rho F_{\nu\rho}D_\mu\right)
\right. \nonumber \\  && \left.  -\beta \left(F_{\mu\rho}D^{\rho}D_\nu-F_{\nu\rho}D^{\rho}D_\mu\right)\right], 
\nonumber \\ &=&  F_{\mu\nu}-2\beta D^\rho D_\rho F_{\mu\nu}
-\beta\left(D^\rho F_{\mu\rho}D_\nu-D^\rho F_{\nu\rho}D_\mu\right)
 \nonumber \\ && -\beta \left(F_{\mu\rho}D^{\rho}D_\nu-F_{\nu\rho}D^{\rho}D_\mu\right)
 \nonumber \\ &=& F_{\mu\nu} + \beta \tilde F_{\mu\nu},  
\label{generalized_tensor}
\end{eqnarray}
 where ${F}_{\mu\nu}=-i[{D}_\mu, {D}_\nu]$ is the un-deformed field tensor, and 
 \begin{eqnarray}
  \tilde F_{\mu\nu} &=&  -2  D^\rho D_\rho F_{\mu\nu}
- \left(D^\rho F_{\mu\rho}D_\nu-D^\rho F_{\nu\rho}D_\mu\right)
 \nonumber \\ && -  \left(F_{\mu\rho}D^{\rho}D_\nu-F_{\nu\rho}D^{\rho}D_\mu\right)
 \end{eqnarray}

 Writing out equation (\ref{covar}) in terms of the undeformed
 potential $A_\mu$, we have
\begin{equation}
\mathcal{D}_\mu = [1-\beta(\partial^\rho +iA_\rho)(\partial_\rho
+iA_\rho)] (\partial_\mu +iA_\mu).
\label{modicovar}
\end{equation}
It is now clear that we can consider a modified potential $\mathcal{A}_\mu$
differing from the undeformed $A_\mu$ by a term proportional to
$\beta$:
\begin{equation}
 \mathcal{A}_\mu= A_\mu + \beta  \tilde{A}_\mu
\end{equation}
with the extra term obtained from equation (\ref{modicovar})
\begin{equation}
\tilde{A}_\mu = (\partial^\rho +iA_\rho)(\partial_\rho
+iA_\rho) (\partial_\mu +iA_\mu).
\end{equation}

We now wish to define loop space variables based on this deformed
field tensor $\mathcal{F}_{\mu\nu}$ to study topological obstructions
in this GUP spacetime.

First we consider the space of loops in undeformed spacetime, with a
fixed base point.
A loop is   parameterized  by the coordinates $\xi^\mu(s)$, 
\begin{equation}
 C : \{ \xi^\mu (s): s = 0 \to 2\pi, \, \, \xi^\mu (0) = \xi^\mu(2\pi)\},  
\end{equation}
where    $\xi^\mu (0) = \xi^\mu(2\pi)$ is the chosen (but arbitrary) 
base point \cite{p1, t1, t2, p}.  Next we define the loop space
variable
\begin{equation}
 \Phi [\xi]  = 
P_s \exp i \int^{2\pi}_0  A^\mu (\xi(s)) \frac{d \xi_\mu}{ds}. 
\end{equation}
where $P_s$ denotes ordering in $s$ 
increasing from right to left.  From this we can define its
logarithmic derivative as  a kind of loop space connection
\begin{equation}
 F_\mu [\xi| s] = i \Phi^{-1}[\xi] \frac{\delta}{\delta \xi^\mu (s)}\Phi[\xi]. 
\end{equation}
The derivative in $s$ is  taken from below. It may be noted as the loop variable $\Phi [\xi]$
only depends on $C$ and not the manner in which $C$ is parameterized, so labeling it with a fixed point is over 
complete.  In fact, any other parameterization of $C$ will  
only change the variable in the integration and not the loop space
variable $\Phi [\xi]$. 

We can obtain a formula relating $ F_\mu [\xi| s]$ to the spacetime
curvature $F_{\mu\nu}$ by first defining a parallel transport 
 from a point $\xi(s_1)$ to a point $\xi (s_2)$ as   \cite{p1, t1, t2, p}
\begin{eqnarray}
 \Phi [\xi: s_1, s_2 ]  =  
P_s \exp i \int^{s_2}_{s_1} A^\mu (\xi(s)) \frac{d \xi_\mu}{ds}. 
\label{transport}
\end{eqnarray}
Thus
\begin{equation}
F^\mu[\xi|s]=\Phi^{-1} [\xi: s,0] F^{\mu\nu} \Phi [\xi: s,0]
\frac{d\xi_\nu(s)}{ds}.
\label{looptolocal}
\end{equation}
This formula can be understood as follows. 
We parallel transport  from a fixed point  along a fixed path  to another fixed point. After reaching that point, 
we will take a  detour then turn 
back along the same path till we reach the  original point. 
Thus, the phase factor generated by 
going along the path from the original point to final point  will be canceled by the phase factor generated by going from the final point back
to the original point.  However, there will be a 
contribution generated by  the transport along the infinitesimal
circuit  along the  final point, which is proportional to the
spacetime curvature at that point.

{ 
We can repeat the same construction using our deformed variables. 
So,  we can define a deformed loop variable with a deformed connection. As this deformed connection, 
is a connection in the deformed theory, we can write  
\begin{equation}
{\mbox{\boldmath $\Phi$}}[\xi] = P_s \exp i \int^{2\pi}_{0} 
\mathcal{A}^\mu (\xi(s)) \frac{d \xi_\mu}{ds}. 
\end{equation}
Here again  $P_s$ denotes ordering in $s$ 
increasing from right to left.
Now we can define  the
logarithmic derivative of this deformed variable  as a deformed loop space connection 
\begin{equation}
 \mathcal{F}_\mu [\xi| s] = i {\mbox{\boldmath $\Phi$}}^{-1}[\xi] \frac{\delta}{\delta \xi^\mu (s)}{\mbox{\boldmath $\Phi$}}[\xi]. 
\end{equation}
We can also deformed a parallel transport 
 from a point $\xi(s_1)$ to a point $\xi (s_2)$ as 
\begin{equation}
{\mbox{\boldmath $\Phi$}}[\xi: s_1, s_2 ]  =  
P_s \exp i \int^{s_2}_{s_1} \mathcal{A}^\mu (\xi(s)) \frac{d
  \xi_\mu}{ds}.
\end{equation}
Now we can use this deformed  parallel transport to go from a fixed point to another fixed point, along a fixed path. 
We can also take a detour from that final fixed point and go back to the initial fixed point along the same path. 
So, the phase generated by going to the final fixed point will exactly cancel the phase generated by going back to 
the initial fixed point. However, to take a detour, we will have to produce an infinitesimal
circuit  along the  final point. This infinitesimal
circuit  will produce a contribution, and as we are using the deformed parallel transport, we can write this 
contribution as  
\begin{equation}
 \mathcal{F}^\mu [\xi|s] =    
{\mbox{\boldmath $\Phi$}}^{-1}[\xi: s,0] \mathcal{F}^{\mu\nu} (\xi (s) )
 {\mbox{\boldmath $\Phi$}}[\xi: s,0]\frac{d \xi_\nu (s) }{d s}.
\end{equation}
Now since the GUP deformation in the covariant derivative is first order
in $\beta$, in this expression we can actually replace the deformed
${\mbox{\boldmath $\Phi$}}$ by the undeformed $\Phi$
\begin{eqnarray}
 \mathcal{F}^\mu [\xi|s] &=&   \Phi^{-1}[\xi: s,0] \mathcal{F}^{\mu\nu} (\xi (s) )
 \Phi[\xi: s,0]\frac{d \xi_\nu (s) }{d s} \nonumber \\ &=& 
 \Phi^{-1}[\xi: s,0] \big[{F}^{\mu\nu} +  \beta \tilde F^{\mu\nu}\big] (\xi (s) )
 \Phi[\xi: s,0]\frac{d \xi_\nu (s) }{d s} \nonumber \\ &=& {F}^\mu [\xi|s] + \beta \tilde{F}^\mu [\xi|s].
\end{eqnarray} 
Now this is important to note that if the original   $  F^{\mu\nu} = 0$, then ${F}^\mu [\xi|s] =0$. 
However, it is possible that  even if $  F^{\mu\nu} = 0$, we can have 
$\tilde F^{\mu\nu} \neq 0$, and so $ \mathcal{F}^{\mu\nu} \neq 0$.
This would mean that even if ${F}^\mu [\xi|s] =0$, we can still have $\tilde {F}^\mu [\xi|s] \neq0$, 
and so $\mathcal{F}^\mu [\xi|s] \neq0$. 
Thus, there could be a contribution to the Polyakov loop  produced
solely from the deformation of the 
background geometry. }

\section{Topological Defects }

We can regard $ \mathcal{F}_\mu  [\xi|s] $ as the connection in the loop space as it 
represents the change in phase of ${\mbox{\boldmath $\Phi$}}[\xi]$ as
one 
moves   in the loop space. 
It is interesting to note that the connection is loop space   $ \mathcal{F}_\mu  [\xi|s]$ is 
proportional to the field strength in spacetime $\mathcal{F}_\mu (\xi(s))$. 
As $ \mathcal{F}_\mu  [\xi|s] $  acts as a 
connection in the loop space, we can define  covariant derivative in loop space  
$\Delta_\mu (s) =  \delta/\delta \xi^\mu (s) + i  \mathcal{F}_\mu  [\xi|s]. 
$ This covariant derivative can  be used to define the  curvature of the loop space $ -i \mathcal{G}_{\mu\nu}[\xi, s_1, s_2]$ 
as  the commutator of these covariant derivatives $
[\Delta_\mu [\xi(s_1)], \Delta_\nu [\xi(s_2)]] $. {   
So, we obtain the following expression for the curvature of the deformed loop space
\begin{eqnarray}
 \mathcal{G}_{\mu\nu}[\xi ( s_1, s_2)] &=&  \frac{\delta}{\delta \xi^\mu (s_2) }\mathcal{F}_\nu  [\xi|s_1]
- \frac{\delta}{\delta \xi^\nu (s_1) }\mathcal{F}_\mu  [\xi|s_2] \nonumber \\&&
+i [\mathcal{F}_\mu  [\xi|s_1], \mathcal{F}_\nu  [\xi|s_2]] \nonumber \\ && 
  {G}_{\mu\nu}[\xi ( s_1, s_2)] + \beta \tilde{G}_{\mu\nu}[\xi ( s_1, s_2)]. 
\end{eqnarray}
Here the original curvature in loop space is given by \cite{t1} 
\begin{eqnarray}
  {G}_{\mu\nu}[\xi ( s_1, s_2)] &=&  \frac{\delta}{\delta \xi^\mu (s_2) } {F}_\nu  [\xi|s_1]
- \frac{\delta}{\delta \xi^\nu (s_1) } {F}_\mu  [\xi|s_2] \nonumber \\&&
+i [ {F}_\mu  [\xi|s_1], {F}_\nu  [\xi|s_2]], 
\end{eqnarray}
and $ \tilde{G}_{\mu\nu}[\xi ( s_1, s_2)]$ is the correction to the original loop space curvature. 
It may  be noted that the deformed loop connection, $ \mathcal{F}_\mu  [\xi|s] $ 
represents a change 
in phase ${\mbox{\boldmath $\Phi$}}$ as one moves 
in the deformed loop space. So, for a deformed gauge connection $\mathcal{A}_\mu$, it is possible 
to construct the holonomy using the deformed field tensor  $\mathcal{F}^{\mu\nu}$. However, now $ \mathcal{F}^\mu [\xi|s]$
is also a connection in the deformed loop space, and so we can construct the corresponding holonomy. Thus, we can go 
from a fixed point in the deformed loop space to another fixed point, and then take a detour back to the initial point. 
We will go back along the same path we initially took, and so the contribution of going to the final point will exactly 
cancel the contribution of going back to the initial point. However, to take a detour, we will have to make an infinitesimal
circuit, and this will have a contribution. 
As we are moving in the deformed loop space, this contribution would be equal to $\mathcal{G}_{\mu\nu}[\xi ( s_1, s_2)]$. 
It may be noted that in spacetime, this would appear as sweeping out an infinitesimal two dimensional surface 
enveloping a three dimensional volume. Now the value of this deformed loop space curvature will depend on what is inside 
this volume. }
This loop space curvature can be used to analyse the presence of a topological defect in the original theory. 
This is because if a monopole is not present in the spacetime, then
this deformed loop space curvature term vanishes. 
This can be seen by   showing that this deformed loop space curvature is
proportional to left-hand side of the Bianchi identity (\ref{bianchi}).

In fact, following closely similar arguments for the usual gauge theories \cite{t1}, we consider variations of the curve in two orthogonal
directions $\lambda$ and $\kappa$.  Now first we define three displaced curves, 
\begin{eqnarray}
(\xi_1^\mu (s))_\lambda &=& (\xi^\mu (s))_\lambda + \Delta \delta^\mu_\lambda \delta
                  (s-s_1) \nonumber\\
(\xi_2^\mu (s))_\kappa &=& (\xi^\mu (s))_\kappa + \Delta' \delta^\mu_\kappa \delta(s-s_2) 
\nonumber\\
(\xi_3^\mu (s))_\kappa &=& (\xi_1^\mu (s))_\kappa + \Delta' \delta^\mu_\kappa \delta (s-s_2),
\end{eqnarray}
where the Kronecker delta $\delta^\mu_\lambda$ means that the
variation is zero if $\mu \neq \lambda$, and similarly for $\delta^\mu_\kappa$.  
Then by definition
\begin{equation}
\frac{\delta}{\delta \xi^\kappa(s_2)} \mathcal{F}_\lambda [\xi|s_1] = \lim_{\Delta
  \to 0} \lim_{\Delta' \to 0} \frac{1}{\Delta \Delta'} \frac{i}{g}
\left\{ \bgphi^{-1}[\xi_2] \bgphi[\xi_3] -
  \bgphi^{-1}[\xi]\bgphi[\xi_1] \right\}. 
\end{equation}
It may be noted that the right-hand side  usually has the implicit indices $\lambda$ and
$\kappa$ as noted above.

Next we calculate  the value of  $ \bgphi^{-1}[\xi_2] \bgphi[\xi_3] 
- \bgphi^{-1}[\xi] \bgphi[\xi_1]$.
Using parallel transport along these paths, we obtain 
\begin{equation}
  \bgphi[\xi_1] = \bgphi[\xi] - i \int ds \bgphi[ \xi: 2\pi, s ]
  \mathcal{F} (\xi(s))  \bgphi(\xi: s, 0), 
\end{equation}  where 
\begin{eqnarray}
  \mathcal{F}  (\xi(s))  &=& 
   \mathcal{F} ^{\mu \nu } (\xi(s)) \frac{d \xi_{ \nu  }(s)}{ds} 
\Delta  \delta_\mu^\lambda \delta(s-s_1).  
\end{eqnarray}
Furthermore, we also obtain, 
\begin{equation}
  \bgphi[\xi_2] = \bgphi[\xi] - i \int ds \bgphi[ \xi: 2\pi, s ]
  \mathcal{F}  (\xi(s))  \bgphi[\xi: s, 0], 
\end{equation}
where 
\begin{eqnarray}
  \mathcal{F}  (\xi(s))  &=& 
   \mathcal{F} ^{\mu \nu } (\xi(s)) \frac{d \xi_{ \nu  }(s)}{ds} 
\Delta' \delta_\mu^\kappa \delta (s-s_2).  
\end{eqnarray}
 Finally, we obtain 
\begin{equation}
  \bgphi[\xi_3] =  \bgphi[\xi_1] - i \int ds \bgphi[ \xi_1: 2\pi, s ]
  \mathcal{F} (\xi_1(s)) \bgphi[\xi_1: s, 0],
\end{equation}
where 
\begin{eqnarray}
  \mathcal{F}  (\xi_1(s))  &=& 
   \mathcal{F} ^{\mu \nu } (\xi_1(s)) \frac{d \xi_{1 \nu  }(s)}{ds} 
\Delta' \delta^\kappa_\mu \delta (s-s_2).
\end{eqnarray}
We can also write  similar expressions for  $\bgphi[\xi: 2\pi, s]$ and 
$\bgphi[\xi_1: s, 0]$.
Now collecting all these these, we obtain the following expression, 
\begin{eqnarray}
 \frac{\delta }{\delta \xi_\mu (s_2)} \mathcal{F}_\nu  [\xi | s_1] &=& 
 \bgphi^{-1}[\xi: s_1, 0]   \mathcal{D}^\nu \mathcal{F}^{\mu\rho} (\xi
 (s_2)) \nonumber \\ 
&& \times\frac{d\xi_\rho (s_1)}{d s_1}\bgphi [\xi: s_1, 0] \delta (s_2 - s_1)\nonumber \\ && + 
 \bgphi^{-1}[\xi: s_2, 0]   \mathcal{F}_{\mu\nu} (\xi (s_2)) 
\bgphi [\xi: s_2, 0]\nonumber \\ && \times \frac{d}{d s_1}\delta (s_2 - s_1)
 \nonumber \\ && +i [\mathcal{F}_\mu  [\xi|s_2], \mathcal{F}_\nu  [\xi|s_1]]\theta (s_1 - s_2). 
\nonumber \\
 \frac{\delta }{\delta \xi_\nu (s_1)} \mathcal{F}_\mu  [\xi | s_2] &=& 
 \bgphi^{-1}[\xi: s_2, 0]   \mathcal{D}^\mu \mathcal{F}^{\nu\tau}(\xi (s_1)) \nonumber \\ && \times \frac{d\xi_\tau (s_2)}{d s_2}\bgphi [\xi: s_2, 0] \delta (s_1 - s_2)
 \nonumber \\ && + 
 \bgphi^{-1}[\xi: s_1, 0]   \mathcal{F}_{\nu\mu} (\xi (s_1)) \bgphi [\xi: s_1, 0]\nonumber \\ && \times \frac{d}{d s_2}\delta (s_1 - s_2)
 \nonumber \\ && +i [\mathcal{F}_\nu  [\xi|s_1], \mathcal{F}_\mu  [\xi|s_2]]\theta (s_2 - s_1). 
\end{eqnarray}
So, the loop space curvature can be written as, 
\begin{eqnarray}
\mathcal{G}_{\mu\nu}[\xi ( s_1, s_2)] &=& \frac{\delta}{\delta \xi^\mu (s_2) }\mathcal{F}_\nu  [\xi|s_1]
- \frac{\delta}{\delta \xi^\nu (s_1) }\mathcal{F}_\mu  [\xi|s_2] \nonumber \\&&
+i [\mathcal{F}_\mu  [\xi|s_1], \mathcal{F}_\nu  [\xi|s_2]]\nonumber \\ &=& 
 \bgphi^{-1}[\xi: s_1,0] \Big[[  \mathcal{D}_\mu ,                                  \mathcal{F}_{\nu\tau}]  +[  \mathcal{D}_\nu , \mathcal{F}_{\tau\mu}] 
 + [  \mathcal{D}_\tau , \mathcal{F}_{\mu\nu}] \Big]
 \nonumber \\ && \times 
 \bgphi[\xi: s_1,0]\frac{d\xi^\tau (s_1)}{ds} \delta (s_1-s_2). 
\end{eqnarray}

Thus,   the deformed  loop space curvature is proportional to the deformed Bianchi identity in the spacetime. 
It is known that the   Bianchi identity are satisfied  in absence of a topological defect in spacetime,
$[  \mathcal{D}_\mu , \mathcal{F}_{\nu\tau}]  +[  \mathcal{D}_\nu , \mathcal{F}_{\tau\mu}] 
 + [  \mathcal{D}_\tau , \mathcal{F}_{\mu\nu}] =0$, 
and so the loop space curvature vanishes in absence of a topological defect in spacetime, 
$\mathcal{G}_{\mu\nu}[\xi ( s_1, s_2)] = 0$. However, if a monopole exists in spacetime, then 
Bianchi identity are not satisfied $[  \mathcal{D}_\mu , \mathcal{F}_{\nu\tau}]  +[  \mathcal{D}_\nu , \mathcal{F}_{\tau\mu}] 
 + [  \mathcal{D}_\tau , \mathcal{F}_{\mu\nu}] \neq0$. Now if the world-line 
of a monopole goes through the point represented by $s_1$, then  the loop space curvature does not 
vanish $\mathcal{G}_{\mu\nu}[\xi ( s_1, s_2)] \neq 0$.  
However, if the topological defect only contributes at the order $\beta$ , then the original Bianchi identity will be satisfied, 
and the deformed Bianchi identity will be violated at the order $\beta$. Thus, the loop space curvature is proportional will also have 
a contribution at the order $\beta$, and it will not vanish. So, it is possible to produce topological defects in the gauge theory from 
the deformation of the background geometry by minimum measurable length scale.
It would be interesting to analyse the consequences of such 
a deformation further. 
\section{Monopole Charge} 
Now we will finally obtain an  expression for the non-abelian monopole charge in such deformed field theories.  
It is possible to obtain the non-abelian monopole charge for the usual gauge theories using the 
concept of loop of loops \cite{t1}. In this section, we will generalize this construction  to deformed gauge theories, 
and thus obtain a generalized monopole charge for deformed gauge theories. 
It is also possible to construct a loop in the loop space by using  
 the connection in the loop space,  $\mathcal{F}^\mu [\xi|s]$. In order to do that, we define  $\Sigma$ as 
\begin{equation}
 \Sigma : \{ \xi^\mu (s): s = 0 \to 2\pi, \, \,  t = 0 \to 2\pi\},  
\end{equation}
where
\begin{eqnarray}
 \xi^\mu (t: 0) = \xi^\mu (t:2\pi), && t = 0 \to 2\pi,  \nonumber \\
 \xi^\mu (0:s) = \xi(2\pi:s), && s = 0 \to 2\pi. 
\end{eqnarray}
So, for each $t$, we have $\xi^\mu(t:s) $ and this  represents a closed loop $C(t)$ $s = 0 \to 2\pi$, 
\begin{equation}
C(t) : \{\xi^\mu (t:s), s = 0 \to 2 \pi \}. 
\end{equation}
Here  $C(t)$ traces out a closed loop as $t$ varies, and 
it  shrinks to a point for  $t =0$ and $ t = 2\pi$.   
Now using $\Sigma$, we can construct a loop in the loop space. 
Thus, for the usual un-deformed gauge theories, this will be given by 
\begin{equation}
 \Theta (\Sigma) =    P_t \exp i \int_0^{2\pi} dt \int^{2\pi}_0 
{F}_\mu [\xi|t,s] \frac{\partial \xi^\mu [\xi|t,s ]}{\partial t}.
\end{equation}
This loop in the loop space is a parameterized surface in spacetime.
Thus, this loop in the loop space encloses a volume. So, it can be used 
to measure the  monopole inside such a volume. We will now 
generalize this construction to deformed gauge theories, and then 
apply that deformed formalism to analyze the monopole charge 
for the deformed gauge theory. 

However, as the deformation by the generalized uncertainty principle,
deforms ${F}_\mu [\xi|t,s]$ to 
$ \mathcal{F}_\mu [\xi|t,s]$, 
we can construct  a loop in the loop space of deformed theories using 
\begin{eqnarray}
 \bgTheta (\Sigma) &=&   P_t \exp i \int_0^{2\pi} dt \int^{2\pi}_0 
 \mathcal{F}_\mu [\xi|t,s] \frac{\partial \xi^\mu [\xi|t,s ]}{\partial t}
 \nonumber\\  &=&  P_t \exp i \int_0^{2\pi} dt \int^{2\pi}_0 
 \big[{F}_\mu [\xi|t,s] + \beta \tilde{F}_\mu [\xi|t,s] \big] \frac{\partial \xi^\mu [\xi|t,s ]}{\partial t}
\nonumber \\ &=& \Theta (\Sigma) + \beta \tilde  \Theta (\Sigma).
\end{eqnarray}
This $\bgTheta$ measures the charge of a non-abelian
monopole for a deformed gauge theory, since $ \bgTheta (\Sigma) =  \zeta$, where 
$\zeta$ is the generalized monopole charge enclosed by the surface
$\Sigma$.   Note that $ \bgTheta (\Sigma) =  I$ the group identity
represents the vacuum, i.e. no topological charge is enclosed by the surface.
As an example, 
let us consider a gauge theory with  $SO(3)$ as its  gauge group. In this case, 
monopole charges are $+1$ for no monopole, and $-1$ for a monopole.
If a monopole is not present, then 
$ \bgTheta (\Sigma)$    will wind fully around the gauge group and will
equal to the identity.
However, in presence of a monopole, 
$ \bgTheta (\Sigma)$    cannot  wind fully around the gauge group and
will equal   the identity.  
It may be noted that the expression  $ \bgTheta (\Sigma) =  \zeta$ is interesting 
as it can be used to evaluate the monopole charge. 
It is possible to demonstrate that this result holds for all non-abelian Yang-Mills theories with  
gauge group is $SU(N)/ Z_N$. The monopole charge for such a gauge group is given 
by $\zeta = \exp i 2 \pi r/N$, where $r = 0, 1, 2, \cdots , (N-1)$. Thus, with this modification 
this result can be applied to Yang-Mills theory with any gauge group.  
It is interesting to note that even the  change has a $\beta$ contribution coming 
from deformation. This occurs because  the topological defects can occur 
at the order $\beta$, even if they do not occur in the original theory. 
Thus, even in deformed gauge theories, the Polyakov loops space formalism can be used to analyse 
the topological defects. 

{  
It may be noted as $\mathcal{G}_{\mu\nu}[\xi ( s_1, s_2)] $ is analogous to $\mathcal{F}^{\mu\nu}$ in deformed 
loop space, it can be constructed using a logarithmic derivative of $  \bgTheta (\Sigma)$. 
So, basically, we can argue that the logarithmic derivative of $  \bgTheta (\Sigma)$ would produce a connection
in this loop of loop space.
In fact, this has been done for ordinary loop space \cite{t1}, and the same argument can be used for 
deformed loop space by using deformed quantities. 
Now for $s_1\neq s_2$, $\mathcal{G}_{\mu\nu}[\xi ( s_1, s_2)] $ does not enclose any volume, 
 and so for this $\bgTheta (\Sigma) = I$, which is the group identity, and its logarithmic derivative vanishes. This also occurs for 
 $s_1 = s_2$, of $\xi(s)$ does not intersect with a monopole worldline, which we can represent by $Y^\rho(\tau)$. So, 
 in that case again $\theta (\Sigma)= I$. However, when $s_1 = s_2$, and $x\i(s) $ intersects a monopole worldline 
 $Y^\rho(\tau)$, $\mathcal{G}_{\mu\nu}[\xi ( s_1, s_2)] $ corresponds to $\Sigma$ enclosing a monopole. 
 Now for original un-deformed loop space variable we have   \cite{t1}, 
\begin{equation}
 {G}_{\mu\nu}[\xi ( s_1, s_2)] = \frac{-\pi}{g} \int d\tau \kappa[\xi|s]  \frac{d \xi_{ \sigma }(s)}{ds} 
  \frac{d Y_{ \rho }(\tau)}{d\tau} 
 \delta(\xi(s) - Y(\tau)) \delta(s_1-s_2), 
\end{equation}
where $\exp\,  i\pi \kappa = \zeta$. However, in deformed loop space, we can write  
$\mathcal{G}_{\mu\nu}[\xi ( s_1, s_2)] = {G}_{\mu\nu}[\xi ( s_1, s_2)] + \beta 
\tilde{G}_{\mu\nu}[\xi ( s_1, s_2)] $, so we obtain 
\begin{eqnarray}
  {G}_{\mu\nu}[\xi ( s_1, s_2)] \nonumber & =&  \frac{-\pi}{g} \int d\tau \kappa[\xi|s]  \frac{d \xi_{ \sigma }(s)}{ds} 
  \frac{d Y_{ \rho }(\tau)}{d\tau} 
 \delta(\xi(s) - Y(\tau)) \delta(s_1-s_2)\nonumber \\ && - \beta 
\tilde{G}_{\mu\nu}[\xi ( s_1, s_2)] . 
\end{eqnarray}
So, even if ${G}_{\mu\nu}[\xi ( s_1, s_2)] =0$, due to the original Bianchi identity being satisfied, we still have 
\begin{equation}
   \frac{-\pi}{g} \int d\tau \kappa[\xi|s]  \frac{d \xi_{ \sigma }(s)}{ds} 
  \frac{d Y_{ \rho }(\tau)}{d\tau} 
 \delta(\xi(s) - Y(\tau)) \delta(s_1-s_2) \neq 0, 
\end{equation}
and so the deformation of the loop space produces a topological defect in spacetime. 
Thus, we will have demonstrated that 
a monopole contribution can  generated from deformation of loop space variables.  
It may be noted that monopoles in general have been analyzed in loop space using a duality which reduces to 
electromagnetic Hodge duality for abelian theories   \cite{pq1,dual,
  dual1,  pq2}. However, 
as far as we know, all such constructions use the loop space formalism,
and we are not aware of any proof of this duality using space-time
variables alone.  Therefore we are restricted at present 
to such a discussion in loop space only. 

We would like to point out that solitonic solutions of the 't Hooft-Polyakov type are sometimes called non-abelian monopoles,
but the magnetic charge carried by them is {  usually} an abelian
magnetic charge 
(with symmetry breaking into a $U(1)$ subgroup).
As far as we know, no solutions of the {pure} Yang-Mills equation  
{  i.e., (without the introduction of symmetry breaking)},
with   a non-abelian monopole charge has been constructed, either using spacetime variables or loop variables.
  Furthermore, the monopoles ({with symmetry breaking})   
  are solutions only outside of a sphere of a finite radius, usually
  interpreted as the 
size of the monopole. 
  Inside of this sphere, not much is known, since the interactions due to the original non-abelian forces become non-negligible.
 Here we are interested in studying the genuninely non-abelian magnetic charge, 
without involving the  Higgs fields.  
Their existence in ordinary spacetime is governed by topology. In spacetimes with a minimum length scale, as we study here,
the obstruction to the vanishing of the relevant loop space curvature indicates also the topological nature 
of this obstruction, which by analogy we think of as generalized monopoles. 
As no non-abelian monopole solutions in ordinary spacetime are known,   to construct one for GUP spacetime
would really be interesting,  but perhaps not feasible at present.  In this paper we 
have demonstrated that minimal length in spacetime can give rise to a  
certain topological charge. However, we would like to point out that it is possible that 
  an object would not exist even if such a charge is allowed to exist \cite{char}. So,  we only demonstrate that such an object can exist 
  due to  the existence of a topological charge produced by minimal length.  }
  
\section{Validity of the Approximation} 
In this section, we will argue that the higher order contributions cannot cancel the topological defects 
produced at a certain order in $\beta$. Thus, we will be able to demonstrate that the
results obtain in this paper are not a consequence of the approximation that we have used. 
This is because if we had considered the deformation to the next order, then 
we would get higher order  contribution to the field strength, which would produce 
higher order contributions to the loop space variables. Thus, if we analyzed the theory to 
the order $\beta^2$, then the corrected field strength would be given by 
\begin{eqnarray}
 \mathcal{F}_{\mu\nu} = F_{\mu\nu} + \beta \tilde F_{\mu\nu} + \beta^2 \bar{\tilde{F}}_{\mu\nu}
\end{eqnarray}
This would occur because $\mathcal{D}_\mu$ will also have a $\beta^2$ contribution to it. This will in 
turn produce a $\beta^2$ contribution to the connection,
\begin{eqnarray}
  \mathcal{A}_\mu= A_\mu + \beta  \tilde{A}_\mu + \beta^2 \bar{\tilde{A}}_\mu. 
\end{eqnarray}
Now by using this new connection in the loop space formalism, we can obtain the $\beta^2$ contribution 
to all the loop space variables.
Thus, by using the expression of the connection to the order $\beta^2$, we obtain 
\begin{eqnarray}
{\mbox{\boldmath $\Phi$}}[\xi] &=& P_s \exp i \int^{2\pi}_{0} 
\mathcal{A}^\mu (\xi(s)) \frac{d \xi_\mu}{ds} \nonumber \\ &=& 
P_s \exp i \int^{2\pi}_{0} 
\big[A^\mu + \beta  \tilde{A}^\mu + \beta^2 \bar{\tilde{A}}^\mu]\big] (\xi(s)) \frac{d \xi_\mu}{ds} ,
\end{eqnarray}
We can also write,  to the $\beta^2$ order,  
\begin{eqnarray}
{\mbox{\boldmath $\Phi$}}[\xi: s_1, s_2 ]  &=&  
P_s \exp i \int^{s_2}_{s_1} \mathcal{A}^\mu (\xi(s)) \frac{d
  \xi_\mu}{ds}
  \nonumber \\ &=& 
  P_s \exp i \int^{s_2}_{s_1} 
  \big[A^\mu + \beta  \tilde{A}^\mu + \beta^2 \bar{\tilde{A}}^\mu]\big] (\xi(s)) \frac{d
  \xi_\mu}{ds}.
\end{eqnarray}
Finally, we can also obtain $ \mathcal{F}^\mu [\xi|s]$ to the order $\beta^2$ as  
\begin{equation}
 \mathcal{F}^\mu [\xi|s] =    
{\mbox{\boldmath $\Phi$}}^{-1}[\xi: s,0] \mathcal{F}^{\mu\nu} (\xi (s) )
 {\mbox{\boldmath $\Phi$}}[\xi: s,0]\frac{d \xi_\nu (s) }{d s}.
\end{equation}
Thus, we can demonstrate that to the order $\beta^2$, 
\begin{eqnarray}
 \mathcal{F}^\mu [\xi|s] &=&   
 \Phi^{-1}[\xi: s,0] \big[{F}^{\mu\nu} +  \beta \tilde F^{\mu\nu} + \beta^2 \bar{\tilde F}_{\mu\nu} \big] (\xi (s) )
 \nonumber \\ &&  \times \Phi[\xi: s,0]\frac{d \xi_\nu (s) }{d s}.
\end{eqnarray}
Thus by repeating this argument we have used for in this paper, to the order $\beta^2$, 
we can demonstrate that to the order $\beta^2$, 
\begin{eqnarray}
\mathcal{G}_{\mu\nu}[\xi ( s_1, s_2)] &=& \frac{\delta}{\delta \xi^\mu (s_2) }\mathcal{F}_\nu  [\xi|s_1]
- \frac{\delta}{\delta \xi^\nu (s_1) }\mathcal{F}_\mu  [\xi|s_2] \nonumber \\&&
+i [\mathcal{F}_\mu  [\xi|s_1], \mathcal{F}_\nu  [\xi|s_2]]\nonumber \\ &=& 
 \bgphi^{-1}[\xi: s_1,0] \Big[[  \mathcal{D}_\mu ,                                  \mathcal{F}_{\nu\tau}]  +[  \mathcal{D}_\nu , \mathcal{F}_{\tau\mu}] 
 + [  \mathcal{D}_\tau , \mathcal{F}_{\mu\nu}] \Big]
 \nonumber \\ && \times 
 \bgphi[\xi: s_1,0]\frac{d\xi^\tau (s_1)}{ds} \delta (s_1-s_2). 
\end{eqnarray}
where we have considered all the covariant derivatives and the field strengths deformed to the order $\beta^2$. 
This is because the $\mathcal{F}^\mu [\xi|s]$ also contain the $\beta^2$ terms. 
It could be demonstrated by repeating the calculations we did to the order $\beta$, that the Bianchi identity
also holds iteratively for higher order $\beta$ deformations, and thus it would hold for $\beta^2$ deformation. 
Thus, we can argue that it would be possible for the 
$\mathcal{G}_{\mu\nu}[\xi ( s_1, s_2)]$ not to be zero at the order $\beta^2$, even if it is zero at the order 
$\beta$. 
Thus, topological defects can occur at higher order, even if they do not occur at lower order. 
However, if $\mathcal{G}_{\mu\nu}[\xi ( s_1, s_2)]\neq 0$ at the order $\beta$, then it cannot vanish at any higher 
order. This is because at higher order say $\beta^2$, the contribution to $\mathcal{G}_{\mu\nu}[\xi ( s_1, s_2)]$ will 
come from $\bar{\tilde{F}}_{\mu\nu}$ which is of order $\beta^2$, and no additional contribution will come at the order 
$\beta$. Now as $\beta<1$, the $\beta^2$ contribution cannot cancel the $\beta$ contributions to the loop space variables. 
Thus, the topological defect which is present at the order $\beta$ cannot be eliminated by considering by considering 
higher order corrections to the loop space. It may be noted that at $\beta^2$ order 
$\bgTheta (\Sigma) = \Theta + \beta \tilde \Theta + \beta^2 \bar{\tilde \Theta} $. 

In fact, this argument can be made iteratively for the loop space variables 
at any order. Thus, if a topological defect exists at the order $\beta^n$,
it cannot be eliminated at the order $\beta^{n +m}$, 
when $m\geq 1$. This is because the $\mathcal{D}_\mu$ will have an contribution to the 
order $\beta^n$ at the order $n$ and $\beta^{n+m}$ at the order $n+m$. 
So, the field strength $\mathcal{F}_{\mu\nu}$ at the order $n$ will also contain terms proportional to $1 \cdots \beta^n$, 
and the field strength $\mathcal{F}_{\mu\nu}$ at the order $n+m$ will contain terms proportional to $1 \cdots \beta^{n+m}$. 
So, the connection $\mathcal{A}_\mu $ will also contain terms 
proportional to $\beta^n$ and $\beta^{n+m}$ at the orders $\beta^n$ and $\beta^{n+m}$, respectively. 
Now  repeating the argument used in this section, we can define the loop space variables 
for each the deformation at any order, and $\mathcal{G}_{\mu\nu}[\xi ( s_1, s_2)]$ would 
also be given in terms of the Bianchi identity at the corresponding  order of the deformation 
parameter.  This implies that  $\mathcal{G}_{\mu\nu}[\xi ( s_1, s_2)]$ will contain 
terms proportional to $\beta^n$ at the order $n$ and $\beta^{n+m}$ at the order $n+m$. 
Now if $\mathcal{G}_{\mu\nu}[\xi ( s_1, s_2)]\neq 0$ 
at $\beta^n$, then this contribution cannot be canceled at the order $\beta^{n +m}$, because 
$\beta <1$. Thus, the topological defects produced at any order cannot be eliminated by considering 
higher order contributions in the deformation parameter. We would also like to point 
out that the $\bgTheta (\Sigma)$ will also contributions proportional to  $1 \cdots \beta^n$
at $n$ order, and $1 \cdots \beta^{n+m}$ at $n+m$ order.

\section{Conclusion}

In this paper, we were able to  analyse the deformation of a gauge theory by the existence of a minimum measurable length scale. This was done using the 
  loop space formalism. We explicitly constructed the loop space variable for this deformed theory. This  loop space variable was then used for 
constructing the loop space curvature. This curvature did not  vanish in presence of a non-abelian monopole. Hence, we were able to demonstrate that 
the non-vanishing of the loop space curvature   
indicates the existence of a topological obstruction
even for deformed gauge theories. We have also constructed an explicit expression for the charge 
of a non-abelian monopole using the loop in the loop space. However, it was possible to consider configurations, for which the original field strength 
vanished, but the deformation did not vanish. Using these field configurations, it was possible to demonstrate that the Polyakov connection can get contributions purely 
from the deformation, and the loop space curvature can also get $\beta$ order contributions,
even if originally it vanished. Thus, it is possible 
the deformation of gauge theories by the deformation of the
background geometry can give rise to topological defects.
We have also demonstrated that higher order corrections 
cannot cancel the topological defects produced at a certain order 
in the loop space formalism. So,  the presence of a minimum length actually may create topological
obstructions like monopoles.  It thus does not
seem to be an artifact of any approximation, but
what may
happen if quantum effects are taken into account, in the
way we propose.  {  
It may be noted that the production of 
magnetic monopoles and even electric charge by
quantum gravitational effects is not a new idea,
and such charges have been constructed using Wheeler-DeWitt approach
\cite{remo, remo1}. 
However, all such work was done only for abelian gauge theories,
and this is the first time  it has been proposed that quantum gravity may produce monopoles  in 
  non-abelian gauge theories. 
We would like to point out that we have only 
 used the deformed gauge theories to obtain such results, however,    such a deformation of  gauge theories occurs due to quantum gravity.  
This is because  a  deformation of quantum mechanics  can occur due to a  
low energy effects from quantum gravity  
\cite{ml12, ml14}, and the   corresponding deformation of quantum field theories (including gauge  theories) can also  occur from 
such quantum gravitational effects 
\cite{qft1, qft2,qft4,qft5,qft6,qft7}. This deformation of  gauge theory, from quantum gravitational effects, is the deformation we have 
used to obtain the results of this paper. 
Thus,  it is possible that the topological defects produced from the deformation studied in this paper, could occurs due to quantum
gravitational effects because of the existence of minimal length in spacetime. 
}

The loop space formalism has been used to construct loop space duality for ordinary Yang-Mills theories  \cite{pq1,dual, dual1,  pq2}.
This duality reduces to the usual electromagnetic Hodge duality for abelian gauge theories. So, even though the Hodge 
duality cannot be generalized to non-abelian gauge theories, this loop space duality can be used to construct a dual 
potential even in case of non-abelian gauge theories. This dual potential has also been used for constructing a 
  Dualized Standard Model   \cite{pq4, p9, pq0, q01, d0}, and which has   in turn been used for   explaining the  
 difference   of masses between   different generations of  fermions  \cite{d1, d2}. 
 This model has also been used for 
  analysing the  
  Neutrino oscillations \cite{no}, 
  Lepton transmutations     \cite{lt}, and    off-diagonal
elements of the CKM matrix  \cite{cmk}. The dual potential used for obtaining the Dualized Standard Model has also been used for 
 constructing 
the 't Hooft's  order-disorder parameters \cite{1t, d, 1p}.  
  It would be interesting to repeat this analysis for a gauge theory deformed by 
a minimum measurable length.
Thus, we can use the results of this paper to construct a dual potential for gauge theories deformed by generalized uncertainty principle. 
This dual potential can in turn be used for constructing a deformed version of Dualized Standard Model. This deformed Dualized Standard Model 
can be used for analyzing the effect on generalized uncertainty principle on  the 
 off-diagonal
elements of the CKM matrix,  difference   of masses between   different generations of  fermions, 
 Neutrino oscillations and  Lepton transmutations. It would also be interesting to analyse the 't Hooft's  order-disorder parameters 
 for gauge theories deformed by generalized uncertainty principle.

\section*{Acknowledgments}

 We would like to thank Mohammed Khalil for   proving the Bianchi identity for deformed gauge theories.

\end{document}